\documentclass[onecolumn,aps,prl,showpacs,amsmath,amssymb,dvipsnames,10pt]{revtex4-2}

\usepackage{graphicx}   
\usepackage{dcolumn}    
\usepackage{bm}         
\usepackage{hyperref}
\usepackage{comment}
\usepackage{graphicx}%
\usepackage{multirow}%
\usepackage{amsmath,amssymb,amsfonts}%
\usepackage{amsthm}%
\usepackage{mathrsfs}%
\usepackage[title]{appendix}%
\usepackage{xcolor}%
\usepackage{textcomp}%
\usepackage{booktabs}%
\usepackage{algorithm}%
\usepackage{algorithmicx}%
\usepackage{algpseudocode}%
\usepackage{listings}%

\usepackage{todonotes}

\usepackage{bm}
\usepackage{ulem}
\def\vec#1{{\ensuremath{\bm{#1}}}}
\newcommand\p[2]{\ensuremath{\frac{\partial #1}{\partial #2}}}
\def\Tr#1{{\ensuremath{\text{Tr}{(#1)}}}}

\def\vec#1{{\ensuremath{{\bm{#1}}}}}

\def\half{{\textstyle \frac{1}{2}}}

\def\p#1#2{{\ensuremath{\frac{\partial #1}{\partial #2}}}}

\begin{document}

\title{Gas-induced bulging in pouch-cell batteries: a mechanical model}

\author{Andrea Giudici}
\email{giudici@maths.ox.ac.uk}
\affiliation{Mathematical Institute, University of Oxford, Woodstock Road, Oxford, OX2 6GG, UK}

\author{Jon Chapman}
\affiliation{Mathematical Institute, University of Oxford, Woodstock Road, Oxford, OX2 6GG, UK}

\author{Colin Please}
\affiliation{Mathematical Institute, University of Oxford, Woodstock Road, Oxford, OX2 6GG, UK}

\date{\today}

\begin{abstract}
  Over the long timescale of many charge/discharge cycles, gas formation can result in large bulging deformations of a Lithium-ion pouch cell, which is  a key failure mechanism in batteries.
  Guided by recent experimental X-ray tomography data of a bulging cell, we propose a homogenised mechanical model to predict the shape of the deformation and the stress distribution analytically. Our model can be included in battery simulation models to capture the effects of mechanical degradation. Furthermore, with knowledge of the bending stiffness of the cathode electrodes and current collectors, and by fitting our model to experimental data, we can predict the internal pressure and the amount of gas in the battery, thus assisting in monitoring the state of health (SOH) of the cell without breaking the sealed case. 
\end{abstract}


\maketitle


\section{Introduction}\label{sec1}

There has been growing interest in understanding how the performance of a lithium-ion battery is influenced by mechanical stresses. In particular, how these stresses affect the charge capacity, internal resistance, and durability of the cell. Usually, stresses arise as the result of the local expansion of electrodes as lithium ions are intercalated into the active material. The expansion can lead to strains as large as $13\%$ (volumetric strain) and, since lithium-ion batteries are composed of layers with different mechanical properties (e.g. electrodes are softer and porous while current collectors are stiffer and non-porous), and the expansion may not be homogeneous across the system, the expansion leads to a complex state of stress. This stress state affects the electro-chemistry and thus the performance of the cell. Simplified models to describe the mechanics and stress state have been developed for spirally wound cells \cite{spirally} and for pouch cells \cite{giudici2024mechanical}. 

Although on the short time-scale of a single charge, lithium-induced swelling is the key ingredient affecting the mechanical state of a battery, on longer timescales --- spanning many charging cycles --- other physical processes, such as gas formation \cite{gas1}, play a crucial role. Gas production is an (often undesirable) outcome of the chemical reactions necessary to operate a battery and tends to increase with the number of charge cycles. In pouch cells, it causes the build up of an internal pressure which leads to deformations, stress, increase in resistance, and a loss of active material, decreasing the overall performance of the battery \cite{Du}.  

Tracking the evolution of gas volume and pressure within a pouch cell would offer a way to monitor the state of health (SOH) of the battery \cite{SOH}. However, measuring the pressure directly requires access to the sealed casing which may not always be practical. An alternative way to measure the amount of gas is by observing the mechanical deformation of the cell, which is characterised by a significant expansion in the through-cell direction. A naive estimate of the pressure can be made by balancing forces on the outer layer of the cell. On the one hand, the gas is trying to expand the system with pressure $P$; on the other hand, the battery material is acting as an elastic spring that opposes the expansion, generating a stress $\propto E \epsilon$ where $\epsilon$ is a measure of the strain and $E$ of the stiffness (Young's modulus) of the material. Balancing the  two, one finds that the strain is $\epsilon \approx P/E$. 

Excluding the very thin separator, which cannot usually accommodate significant strains, the materials that make up a battery tend to have a reported stiffness of about $1$ GPa \cite{uccel2022,guptacathode}, suggesting that even for small strains of order of $1\%$, the internal pressure within the battery needs to be as large as $10^2$ atmospheres --- an exceedingly large value. Indeed, although there has not been much experimental work on directly measuring gas pressure, some studies report pressures which are about two orders of magnitude smaller \cite{pressure1}, on the order of one atmospheric pressure. 

Our naive analysis suggests that the only way in which we can get a significant strain in the through-cell direction with more realistic pressures is if the effective stiffness of the battery under extension is lower than the reported values. This softening under tension of the anode with modest strains has already been reported in \cite{uccel2022}, and is presumably exacerbated as the SOH of the battery decreases. Unfortunately, our naive analysis can only offer a way to measure the relative importance of pressure compared to stiffness, but not their absolute magnitude. Since measuring the effective stiffness of a battery in operando is not trivial; a better approach is required.

\begin{figure*}[t]
    \centering
    \includegraphics[width=\textwidth]{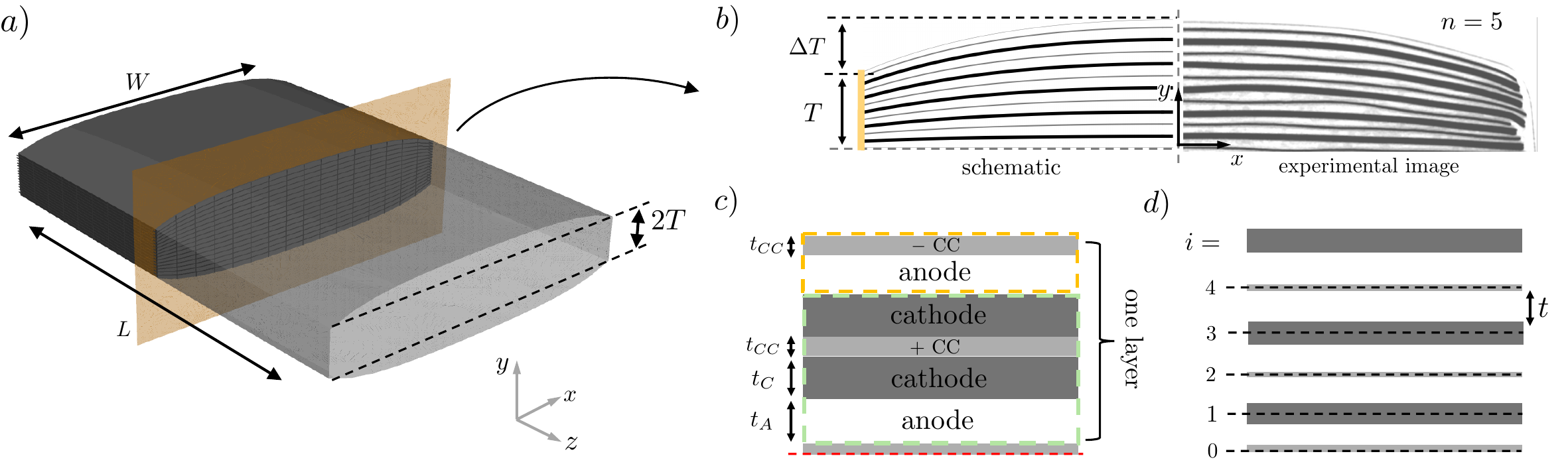}
    \caption{ a) A 3D schematic of the deformation in the pouch cell due to gas pressure. b) Side by side of a schematic view of the cell deformation (left) and the experimental X-ray tomography data from \cite{Du} taken half way along the cell (right). c) Schematic of the battery sublayers. In dashed green, the  anode with the cathode plus positive CC sublayer. In dashed yellow, the anode plus negative CC sublayer. The two sublayers together form one battery layer. d) Simplified version of the battery with anodes in white, CC bending layers in light gray and cathode bending layers in dark gray. The distance between the mid-planes of two consecutive bending layers (dashed black line passing through the midplane of current collectors) is $t$ and is half the thickness of one battery layer.}
    \label{fig:intro}
\end{figure*}

Fortunately, recent X-ray tomography data of a pouch cell published in \cite{Du} allows us to learn a great deal about the gas-induced deformation in a pouch cell. The detailed images show a large and complex bulging deformation, with strains as large as $70\%$ only in the middle of the structure, while the edges of the structure remain clamped by the casing. A schematic of the deformation is shown in figure \ref{fig:intro}a, while in figure \ref{fig:intro}b shows an experimental X-ray tomography image of a slice mid-way along the battery taken from \cite{Du}.
The data clarifies that most of the through-cell deformation occurs in the anode layers which undergo a significant change in thickness. Conversely, the thickness of the cathode layers remain mostly unchanged and, together with their respective current collectors, they appear to act as bending sheets, opposing large curvature and thus determining the shape of the bulge. This new level of detail offers the unique opportunity to model the behaviour of the system and use it to predict the pressure (and effective stiffness) of the system, as well as simulating the degradation process during cycling. 

In this paper, we propose a simple mechanical model that describes the shape of bulging deformation in a pouch cell as a function of the internal pressure in the system, the bending stiffness of the cathode layers and current collectors, the stiffness of the anode layers, and the internal pressure. In the first part of the paper, we lay the assumptions of our model: the system is composed of a set of alternating bending sheets --- the cathode layers and the negative CCs --- connected by softer springy (Winkler) foundations --- the anodes. Inside each layer the gas pressure  expands the system, favouring an isotropic increase in size of the battery. However, the inextensibility of the stiff current collectors and the clamping introduced by the casing causes the system to bulge and expand predominantly in the centre. 

To outline the key ingredients in the problem, we first discuss the simple example of a battery composed of one layer. We show that this simple problem is characterised by two dimensionless numbers, one associated with the strain induced by the pressure and the other with the typical length-scale over which a bending deformation persists from the clamps to the bulk --- a result of the competition between bending and stretching. 
We then proceed to study the full system: we write down the energy for a general battery with several layers. To make analytical progress, we exploit the small aspect ratio of the electrodes, which are thin and long, and write a homogenised theory that describes the shape of the deformation in the limit in which the number of layers becomes large. Once again, by applying the appropriate rescalings, we show that two dimensionless parameters, associated with strain and the persistence length of bending, govern the behaviour of the system. We fit our model to the experimental data in \cite{Du} to predict the pressure and stiffness in the pouch cell. We obtain a realistic estimate for the pressure: of the order of one atmospheric pressure, consistent with range of previously reported values \cite{pressure1}. Finally, we show that our model can be used to predict the amount of gas in the battery, which grows super-linearly as a function of the pressure. This behaviour depends on the persistence length of bending in the cell and offers a novel way to estimate the SOH of the battery simply via measurement of the bulging deformation.

\section{Mechanical model}

We consider the pouch cell from the experimental study in \cite{Du}. The cell has reported length $L=49$mm, width $W=22.5$mm and half-thickness $T=2$mm.  Note that here we have chosen the $x$-axis to be the in-plane direction. This is in line with much work on mechanical deformations of slender objects (i.e. beams), but in contrast with literature on batteries, in which $x$ is commonly the through-cell direction ($y$ in this work). 

The battery is a layered structure made of anode electrodes, cathodes, their respective current collectors and thin separators enclosed in a casing. Since the separator is soft and thin, we will include its effects in the anode mechanics. As gas is produced within the battery, the structure tries to expand in all directions. However, since the current collectors are very stiff and thus inextensible, no expansion in the $x$ direction is observed. Furthermore, the structure is effectively clamped at the ends by the casing, meaning no displacement is allowed in this region (the thickness at the edges has less than $3\%$ variation after 200 cycles). As a result of these constraints, the cell bulges and becomes thicker only in the middle, as shown in figure \ref{fig:intro}.

The experimental data suggests that the deformation is broadly the same along the length ($z$ axis) apart from a relatively small region near the ends (at $z=\pm L/2$). This is because the length is larger than the width, $L>W$, meaning that in the bulk the deformation can be modelled as a 2D plane strain deformation in the $x$-$y$ plane. 
The images also suggest that most of the displacement occurs in the anode layers --- where the gas collects forming pockets --- with the total thickness appearing to change by more than $50\%$ after 200 cycle. 
In contrast, the cathode layers hardly change in thickness, with a total reported change of about $3\%$. This suggests that the anodes are much softer than the cathodes and are the only component of the battery that undergoes large through-cell strain, while cathodes and their current collectors only contribute with their bending rigidity.  

These observations motivate modelling the battery as a 2D structure composed of a stack of two different types of sublayers. Each sublayer is composed of a soft substrate that behaves like a Winkler foundation, the anode electrode, and a stiff bending sheet. The bending sheets come in two flavours: a negative current collector or a cathode layer with its positive current collector, as shown in figure \ref{fig:intro}d. We refer to them as the CC and cathode sublayers, respectively. A battery layer is composed of two sublayers, one of each type.

Note that, although cathode electrodes are about two orders of magnitude softer than CCs, they are about one order of magnitude thicker. Since the bending stiffness of a sheet scales with the cube of the thickness but it is linear in the stiffness, we expect the cathode plus positive CCs to behave as a significantly stiffer bending sheet when compared to the negative CC alone. Thus, we expect the former to dominate the mechanics in a typical battery.

We assume that cathodes, CCs and anodes are made of isotropic linear elastic materials with stiffnesses $E_{CC}$, $E_C$, $E_A$ and Poisson ratio $\nu_{CC}$, $\nu_C$ and $\nu_A$ respectively. Since we assume that both the cathode layers and the CCs do not undergo any change in thickness during the deformation, it is convenient to measure displacements of the bending layers from the mid-plane of the current-collectors, as show in figure \ref{fig:intro}d. Crucially, the distance between consecutive mid-planes in the reference (pristine) cell is always $t=t_C+t_A+t_{CC}$, where $t_C$ is the thickness of the cathode electrode, $t_A$ is the thickness of the anode electrode, and $t_{CC}$ is the thickness of the CCs (we assume, for simplicity, that both current collectors have the same thickness). This means that the total thickness of the battery in the reference (or pristine) state with $n$ layers is $T=2 n t$.

Inside the battery, the build up of gas generates a pressure $P$ which expands the system. This expansion leads the mid-plane of the $i^{th}$ bending layer to deform, with a vertical displacement given by $v_i(x)$. Since the deformation is constrained at the ends by the casing, we also assume that at $x=\pm W/2$ there is no displacement and $v_i=0$. A schematic of the deformation and the sublayers are shown in figure \ref{fig:intro}b) and c).

To study the deformation of the system, we write down the total potential energy associated with the deformation and minimise it with respect to variations in the displacements $v_i$ subject to clamped boundary conditions. However, before studying the full system with $n$ stacked layers, we begin by treating a simpler problem: we consider the deformation of a system composed of only one bending sheet and a soft substrate.

\subsection{Single layer}

We begin by considering a system composed of only one sublayer: a bending sheet attached to a soft foundation bulging due to internal pressure. For now, we shall not worry about what the bending sheet is made of, i.e. a cathode layer or just a CC. We shall solve the general problem and assume that the bending sheet is inextensible and has bending stiffness $\bar{B}$. Conversely, we may treat the soft layer as the anode since this is always the same in each sublayer. 

The total energy stored in the system is the sum of the energy cost associated with bending the sheet, that of stretching the springy substrate of the anode, and the energy saved by pressure as the system expands. We assume the displacement of the bending sheet $v_1$ is small compared to the length of the system, meaning  $v_1/ W \ll 1$ and that the gradients of the deformation are small $\p{v_1}{x} \ll 1$. Then, for small bending deformations (i.e. the radius of curvature is large compared to the thickness of the layer) the curvature can be written as $\kappa_1=\frac{\partial^2 v_1}{\partial x^2}$ and the bending energy is
\[
U_{B}= \int_{-W/2}^{W/2}\frac{L \bar{B}}{2} \left(\frac{\partial^2 v_1}{\partial x^2}\right)^2 dx.
\]

The energy stored in the anode is the energy stored in the springs of stiffness $K$ that make up the Winkler foundation. The restoring force in the anode layer when strained is  $F = Kv_1/t_A $.
Therefore, the energy in each anode layer is
\[
U_{el}=\int_{-W/2}^{W/2}\frac{L K}{2}\frac{v_1^2}{t_A} dx.
\]
In appendix A, we show that when the anode layer is thin and the gradients $v_i$ remain small, the elastic anode layer behaves like a Winkler foundation with $K={ E_A (1-\nu_A)}/{(1+\nu_A)(1-2 \nu_A)} $.

Finally, we assume that the gas moves freely through the cell so that its pressure is spatially uniform.
Due to the inextensibility of the bending layer, vertical displacements $v_1$ lead to horizontal displacements of order $(v_1/W)^2 \ll 1$. However, since these are small, we only consider the energetic contribution of vertical expansion so that the energy saved by expanding the structure is
\[ U_P=- L \int_{-W/2}^{W/2} P v_1\, dx.\] 
The total energy of the system is thus given by:
    \begin{equation}
    \label{eqn:energy_tot_1}
    U_1= L \int_{-W/2}^{W/2}\left( \frac{K}{2} \frac{v_{1}^2}{t_A} +\frac{\bar{B}}{2} \left(\frac{\partial^2 v_1}{\partial x^2}\right)^2 - P v_1\right)dx.
    \end{equation}
We can minimise this energy with respect to variations in $v_1$, leading to the equilibrium equation:
\begin{equation}
\label{eq:eq1D}
   \bar{B} \frac{\partial^4 v_1}{\partial x^4}+ \frac{K}{t_A}v_1-P=0.
\end{equation}
As boundary conditions, we impose clamped ends and no bending moment at the sides, so that $v_1=0$ and $\frac{\partial^2 v_1}{\partial x^2}=0$ at $x=\pm W/2$.

When the bending stiffness is small, we expect the pressure and stress in the elastic substrate to balance, leading to a total strain $\epsilon_1=P/K$. We use this to non-dimensionalise the problem and we write:
\begin{equation}
v_1=\epsilon_1 t_A  (1+X_1),  \quad   x = W \bar{x}.
 \end{equation}
So that the equilibrium condition \eqref{eq:eq1D} becomes:
\begin{equation} 
\label{eq:V}
\frac{\partial^4 X_1}{\partial x^4} + 4 \gamma_1^4 X_1=0,
\end{equation}
where the dimensionless parameter
\begin{equation}
\gamma_1=\left(\frac{W}{l_1}\right)
\end{equation}
determines the ratio between the width $W$ of the system and  the typical length-scale that arises as a result of the competition between bending stiffness of the sheet and stretching of the substrate  
\begin{equation}
    l_1= \sqrt[4]{4 \bar{B} t_A/K}.
\end{equation}
This is the length-scale over which a deformation at the edge of the surface persists into the bulk.

The boundary conditions are:

\begin{equation}
\label{eq:l1}
    X_1=-1, \quad \frac{\partial^2 X_1}{\partial x^2}=0, \quad \text{at $x=\pm W/2$}.
\end{equation} 

Equation \eqref{eq:V} with boundary condition \eqref{eq:l1} is the same equation that governs the behaviour of an elastic beam supported on a soft foundation indented at one side \cite{landau2020theory} and has solution
\begin{equation}
    X_1(\bar{x}) = a \cosh \left(\gamma_1 \bar{x}\right) \cos\left(\gamma_1 \bar{x}\right) +   b \sinh \left(\gamma_1 \bar{x}\right) \sin\left(\gamma_1 \bar{x}\right),
\end{equation}
where the boundary conditions require
\begin{align}
    a=b \cot \left( \half \gamma_1 \right) \coth\left( \half \gamma_1 \right) \quad \text{and}\quad 
    b=\frac{2 \sin \left(\half \gamma_1\right) \sinh \left(\half \gamma_1\right)}{\cos
   \left( \gamma_1\right)+\cosh \left(\gamma_1\right)}.
\end{align}

\begin{figure}[th]
    \centering
    \includegraphics[width=0.6\textwidth]{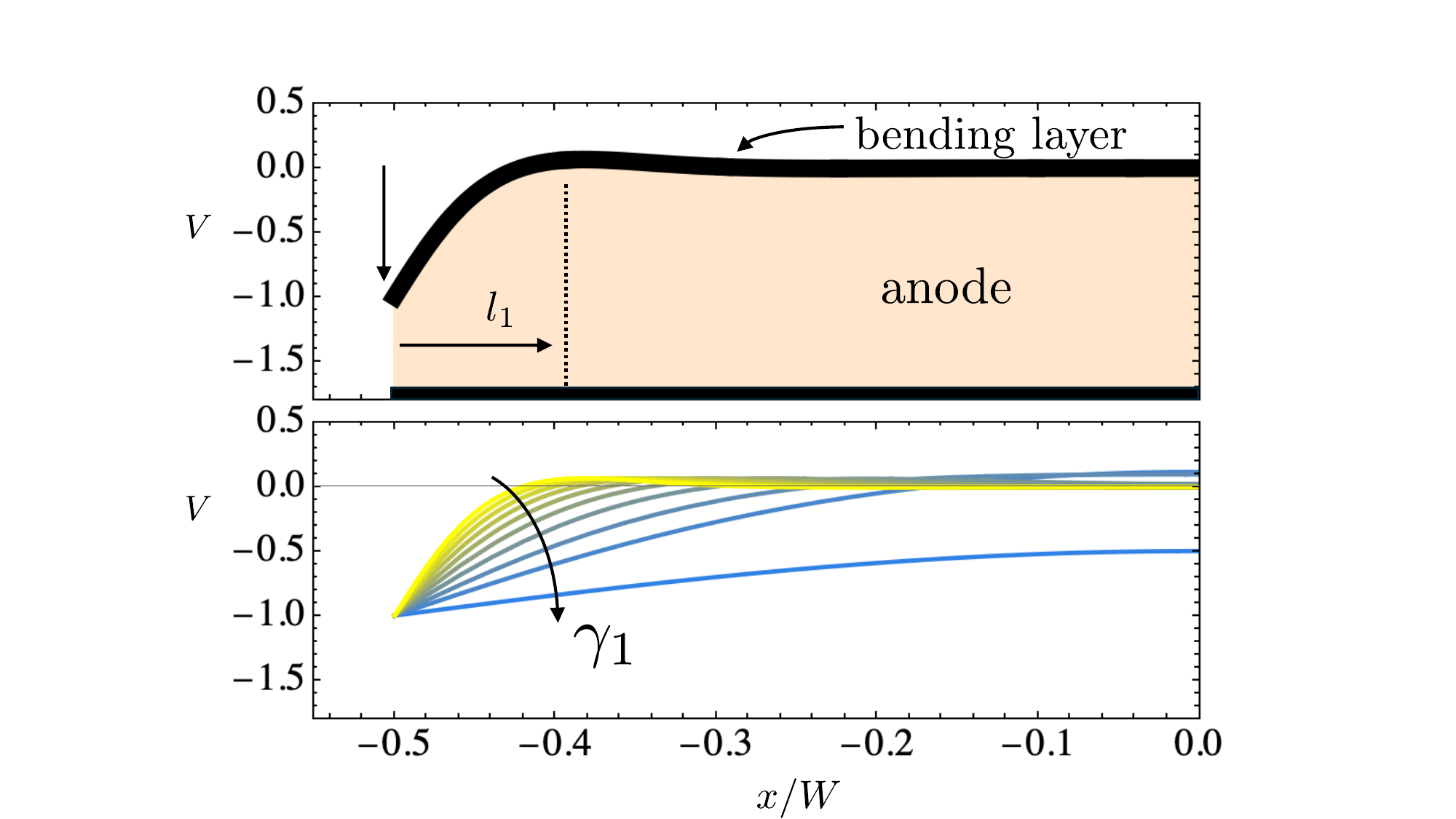}
    \caption{ At the top, we show the shape of a sheet attached to a soft substrate subject to no displacement boundary condition, from $-0.5$ to $0$. We show one side only, since the other is identical due to symmetry. In the second image we show how the key quantity determining the shape of the system is the ratio $\gamma_1=l/W$, with increasing values leading to less localised boundary effects.  }
    \label{fig:1D}
\end{figure}

This simpler problem highlights how the shape of the system is entirely characterised by two parameters: the ratio between the typical length-scale and the width of the domain, $\gamma_1=W/l$, and the strain induced  by the pressure, $\epsilon_1=P/K$, as shown in figure \ref{fig:1D}.

\subsection{Multiple layers}
We are now ready to consider the full problem and study the deformation that occurs when multiple sublayers are stacked together, forming a battery. Without loss of generality, and reflecting the experimental images, we shall assume that the first layer being stacked from the symmetry line is a cathode layer, as shown in figure \ref{fig:intro}d). 
Our approach is to write down the energy of the system for $n$ layers ($2 n$ sublayers) and minimise it with respect to variations in the displacement of each sublayer. The bending stiffness of the current collector is $B_{CC}=\frac{E_C t_{CC}^3}{12(1-\nu_{CC}^2)}$ while, the bending stiffness of the cathode plus positive CC is $B_C=B_{CC}-\frac{E_{C} t_{CC}^3}{12(1-\nu_{C}^2)}  + \frac{2 E_{C}}{3(1-\nu_{CC}^2)} (\half t_{CC}+t_{C})^3$ since we need to consider the contribution from both materials (details on how to calculate the bending stiffness in Appendix B). Using these results, we may write the total energy of the system as:
 \begin{widetext}
    \begin{align}
    \label{eqn:energy_tot_n}
    \tilde{U}_{tot}= L \int_{-W/2}^{W/2}\left[\sum_{i=1}^{2 n}\frac{K}{2} \frac{(v_{i}-v_{i-1})^2}{t_A} +\sum_{i=1}^{n}\frac{B_C}{2} \left(\frac{\partial^2 v_{2i-1}}{\partial x^2}\right)^2+\sum_{i=1}^{n}\frac{B_{CC}}{2} \left(\frac{\partial^2 v_{2i}}{\partial x^2}\right)^2 -\sum_{i=1}^{2 n} P(v_i-v_{i-1})\right]dx,
    \end{align}
     \end{widetext}
where the first term is the stretching energy of the anodes, the second and third terms are the bending energy of the cathodes and current collectors while the latter is the energy saved by the pressure in expanding (each) sublayer.

We can minimise this energy with respect to variations in the displacements $v_i(x)$, leading to $2n$ coupled differential equations which we need to solve for given $K$, $B_C$, $B_{CC}$, $P$ and the appropriate boundary conditions. Such system cannot be readily solved analytically. We will later show numerical solutions to the discrete problem and compare them with the analytical results (details on the numerical simulations are discussed in Appendix C.) To obtain an approximate analytical solution and simplify our problem, we look at the limit in which there are many thin layers, allowing us to homogenise the system and solve it exactly. 

\subsection{Homogenised model}

 Batteries are composed of roughly 10 cathode and 10 anode electrodes ($n=5$) which are about $80$ to $90$ $\mu m$ thick and a couple of centimetres wide and long, meaning the layers are very thin. Therefore, we take the physically relevant limit in which the thickness $t=T/(2n)$ becomes small and the number of layers $n$ becomes large.
If we define
 \begin{equation}
     \bar{B}=\half(B_C+B_{CC})\quad \text{and} \quad \Delta{B}=\half(B_C-B_{CC}).
 \end{equation}
and let $t_A=\phi t$ where $\phi \in [0,1] $, 
we may re-write equation \eqref{eqn:energy_tot_n} as 
\begin{equation}
    \label{eq:discrete_energy}
    \tilde{U}_{tot}=L\int_{-W/2}^{W/2}\sum_{i=1}^{2 n}\left( \frac{K}{2\phi} \left(\frac{(v_{i}-v_{i-1})}{t}\right)^2 + \frac{1}{2t}\left(\bar{B}+(-1)^{i-1} \Delta B \right) \left(\frac{\partial^2 v_i}{\partial x^2}\right)^2  - P\frac{(v_i-v_{i-1})}{t}\right)t \, dx.
\end{equation}
Then, taking the limit as $t \to 0$ (equivalent to $n \to\infty$) and identifying $t$ as $dy$, we obtain a continuous version of the energy:
\begin{equation}
    \tilde{U}_{tot}=L\int_{-W/2}^{W/2} \int_{0}^{T}  \left[\frac{\hat{K}  }{2} \left(\p{v}{y}\right)^2 + \frac{\hat{B}}{2}   \left(\frac{\partial^2 v}{\partial x^2}\right)^2-P \left(\p{v}{y}\right) \right] dx \,dy.
    \label{eq:energyContinuousB}
\end{equation}
where $\hat{K}=K/\phi$ and $\hat{B}=\bar{B}/t$. In the many thin-layers limit, the displacement field $v_i(x)$ has become continuous along the $y$ coordinate and is now a field that depends on both $x$ and $y$, $v(x,y)$. Note that the  alternating term in the energy in equation \eqref{eq:discrete_energy} vanishes as $n \to \infty$ by the alternating series test since $|\Delta B \frac{\partial^2 v}{\partial x^2} |$ is bounded. 

Having a continuous version of the energy means we can now use standard calculus of variations and find the shape of the deformation by solving a partial differential equation for the displacement $v$. The PDE that governs the shape of the system can be found by minimising the energy in \eqref{eq:energyContinuousB} with respect to variations in $v(x,y)$, leading to the bulk equilibrium equation:
\begin{equation}
\label{eq:equilibrium}
    K \frac{\partial^2 v}{\partial y^2}- \hat{B} \frac{\partial^4 v}{\partial x^4}=0.
\end{equation}
with boundary conditions: 
\begin{align}
\label{BCA1}
v&=0 \quad   \text{at }y=0,\\
\label{BCA2}
v&=0 \quad   \text{at }x=\pm L/2,\\
\label{BCA3}
\frac{\partial^2 v}{\partial y^2}&=0 \quad   \text{at }x=\pm L/2,\\
\label{BCA4}
K \frac{\partial v}{\partial y}-P + B_{case}\frac{\partial^4 v}{\partial x^4}&=0 \quad   \text{at }y=T, 
\end{align}
where \eqref{BCA1} and \eqref{BCA2} apply the symmetry about the $x$ axis and the clamping at the edges caused by contact with the casing respectively. Equations \eqref{BCA3} and \eqref{BCA4} tell us that there is no curvature at the edges, where the beams come in contact with the casing, and that vertical forces balance out on the outer layer of the structure. In the latter condition, we have also included the term that would arise if one also considered the bending stiffness of the case, $B_{case}$, to play an important role in the problem. However, here, we assume it is negligible compared to the stiffness of the cathode layers, and therefore ignore the term hereafter. 

\subsubsection{Non-dimensional equations:}

Following the rescaling for the one-layer system, we non-dimensionalise the problem using the strain induced by the pressure $\epsilon_n={P}/{\hat{K}}$, and rescale lengths so that
\begin{equation}
v=\epsilon_n T \bar{v},  \quad   x = W \bar{x},  \quad   y = T \bar{y}.
 \end{equation}
The bulk equation becomes
\begin{equation}
4 \gamma^4\frac{\partial^2 \bar{v}}{\partial \bar{y}^2}-  \frac{\partial^4 \bar{v}}{\partial \bar{x}^4}=0,
    \label{deqRescaled}
\end{equation}
with 
\begin{align}
\label{eq:gamman}
\gamma = \left(\frac{\hat{K} W^4}{4\hat{B}T^2}\right)^{1/4}=\left(\frac{W}{l_n}\right).
\end{align}
The new length-scale is given by $l_n = \sqrt[4]{16 \bar{B} n^2 t \phi / K}=\sqrt{2 n} l_1$ where $l_1$ is the length-scale for the one-layer problem defined in equation \eqref{eq:l1}. This means that the lengthscale associated with the decay of a deformation on the surface of the layered structure is proportional to the lengthscale that arises for a single layer and the square root of the number of sub-layers.
Finally, the boundary conditions are:
\begin{align}
\label{bca}
\bar{v}&=0 \quad   \text{at }\bar{y}=0,\\
\label{bcb}
\bar{v}&=0 \quad   \text{at }\bar{x}=\pm \half,\\
\label{bcc}
\frac{ \partial^2 \bar{v}}{\partial \bar{x}^2}&=0 \quad   \text{at }\bar{x}=\pm \half,\\
\label{bcd}
\frac{\partial \bar{v}}{\partial \bar{y}}-1 &=0 \quad   \text{at }\bar{y}=1.
\end{align}
The boundary value problem in equation \eqref{deqRescaled} with boundary conditions \eqref{bca},\eqref{bcb},\eqref{bcc} and \eqref{bcd} has a separable solutions of the form
\begin{equation}
\label{eq:vsol}
    \bar{v}=\bar{y}+\sin(\lambda \bar{y}) X(\bar{x}),
\end{equation}
where $X$ satisfies
\begin{equation}
    X''''+4 
    \lambda^2\gamma^4 X=0, 
\end{equation}
which has exactly the same form of equation \eqref{eq:equilibrium}. It has the symmetric solution:
\begin{equation}
    X(\bar{x})=A \cosh(\sqrt{\lambda}\gamma \bar{x})\cos(\sqrt{\lambda} \gamma \bar{x})+B \sinh(\sqrt{\lambda}\gamma\bar{x})\sin(\sqrt{\lambda} \gamma \bar{x}).
\end{equation}

Now, boundary condition \eqref{bca} is already satisfied since we chose $v$ to be odd in $y$, eqn \eqref{eq:vsol}. The second boundary condition, equation \eqref{bcd}, is satisfied providing 
\begin{equation}
    \lambda_m = \frac{(2 m+1)\pi}{2}, \qquad m \in \mathbb{N}.
\end{equation}
Thus, the solution can be written as:
\[
 \bar{v}(\bar{x})=\bar{y} + \sum_{m=1}^{\infty} A_m \cosh(\sqrt{\lambda_m} \gamma \bar{x})\cos(\sqrt{\lambda_m} \gamma \bar{x})\sin(\lambda_m \bar{y})
    +\sum_{m=1}^{\infty} B_m \sinh(\sqrt{\lambda_m}\gamma \bar{x})\sin(\sqrt{\lambda_m} \gamma \bar{x})\sin(\lambda_m \bar{y}).
\]
The vanishing curvature boundary condition, equation \eqref{bcc}, becomes
\begin{equation}
    B_m \cosh(\half \sqrt{\lambda_m}\gamma\,)\cos(\half \sqrt{\lambda_m}\gamma\,)\sin(\lambda_m \bar{y})=A_m \sinh(\half \sqrt{\lambda_m}\gamma\,)\sin(\half \sqrt{\lambda_m}\gamma\,)\sin(\lambda_m \bar{y})
\end{equation}
and is satisfied when
\begin{equation}
    B_m= A_m \tanh(\half \sqrt{\lambda_m} \gamma \,)\tan(\half \sqrt{\lambda_m} \gamma\,).
\end{equation}
\begin{figure*}[t]
    \centering
    \includegraphics[width=\textwidth]{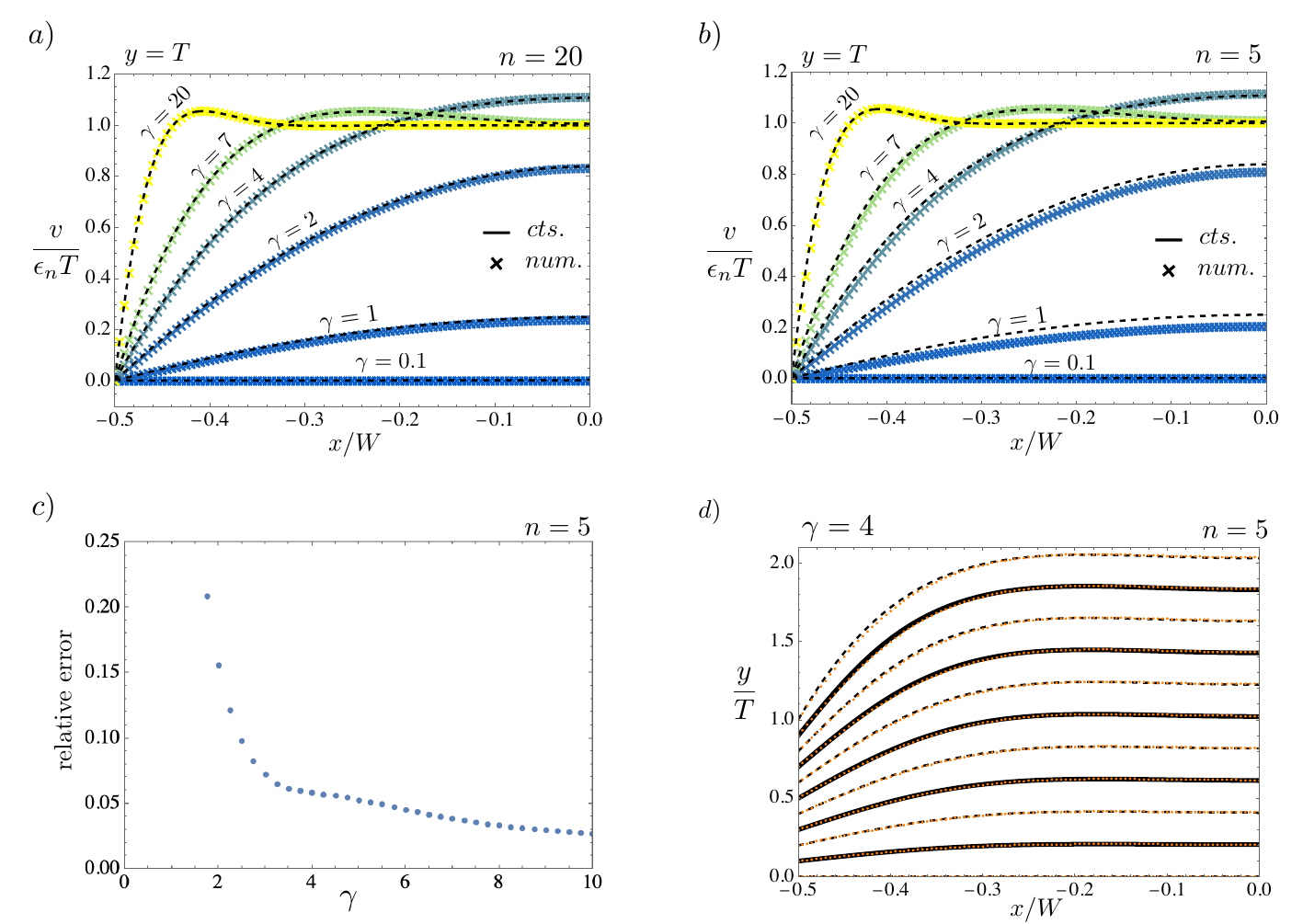}
    \caption{Comparison of the shape of the top layer between the continuous (homogenised) model and the (numerical) discrete model for batteries with different number of layers, (a)  $n=20$ and (b) $n=5$, at different values of $\gamma$. c) The relative error between the continuous and discrete models  at $n=5$. The error is less than $6\%$ for $\gamma>3$. d) Comparison between continuous and discrete models of all the $n=5$ layers when $\gamma=4$. The dashed lines are the negative CCs while full lines the cathode bending layers. In all the plots, $\Delta B/\bar{B} = 0.1$. }
    \label{fig:NDmodel}
\end{figure*}
Finally, boundary condition \eqref{bcb} requires
\begin{widetext}
\begin{equation}
\bar{y}+\half\sum_{m=1}^{\infty} A_m \sec \left(\half  \sqrt{\lambda_m}\gamma \,\right)
   \text{sech}\left(\half \sqrt{\lambda_m}\gamma
  \right) \sin (\lambda_m 
   \bar{y}) \left(\cos \left(\sqrt{\lambda_m}\gamma\right)+\cosh \left(\sqrt{\lambda_m}\gamma
   \right)\right)=0.
\end{equation} 
\end{widetext}
We can solve this condition using a Fourier series. Multiplying  by $\sin (\lambda_k \bar{y})$ and integrating in the symmetric domain $\bar{y} \in [-1,1]$ we obtain:
\begin{equation}
    A_m=- \frac{16 (-1)^{m} \cosh(\half \sqrt{\lambda_m}\gamma\,)\cos(\half \sqrt{\lambda_m}\gamma\,)}{\pi^2 (2m+1)^2  \left(\cosh( \sqrt{\lambda_m}\gamma\,)+\cos( \sqrt{\lambda_m}\gamma\,) \right)}
\end{equation}
which reassuringly converges like $1/m^2$. The final solution reads:
\begin{widetext}
\begin{multline}
\label{eq:vbar}
    \bar{v}(\bar{x},\bar{y}) =  \bar{y}-\sum_{m=0}^{\infty} A_m \sin (\lambda_m \bar{y} ) \left(\cos \left(\bar{x} \sqrt{\lambda_m}\gamma\right) \cosh \left(\bar{x}
   \sqrt{\lambda_m}\gamma\right)\right)\\
   -\sum_{m=0}^{\infty} A_m \sin (\lambda_m \bar{y} ) \left(\tan \left(\half \sqrt{\lambda_m}\gamma\right) \tanh \left(\half \sqrt{\lambda_m}\gamma\right) \sin
   \left(\bar{x} \sqrt{\lambda_m}\gamma \right) \sinh \left(\bar{x}
  \sqrt{\lambda_m}\gamma \right)\right).
\end{multline}
\end{widetext}
Just as in the case of the single layer system, the shape of the deformation is entirely determined by the parameter $\gamma$, while the amplitude of the deformation is determined by $\epsilon_n$. In figure \eqref{fig:NDmodel} we show the shape of the bulge predicted by the homogenised model, equation \eqref{eq:vbar}, at different vales of $\gamma$ and compare it with numerical minimisation of the discreete model given by equation \eqref{eq:discrete_energy} (details found in Appendix B). The agreement is excellent for all values of $\gamma$ when $n=20$ (a battery with 40 layers), panel (a). For $n=5$ (a battery with 10 layers), the case of our battery, the agreement is good when $\gamma>3$ (less than $6\%$ relative error), but worsens when $\gamma<3$ yielding an error of more than $20\%$ for $\gamma<2$.
As we shall see, the experiments are in the region well approximated by our continuous model, and the continuous theory can therefore be used to estimate the parameters of the system. We therefore use the continuous model, and check {\em a posteriori} that the parameters lie in its region of validity.

\section{Comparison with Experiments and prediction of the pressure}

\begin{figure*}[t]
    \centering \includegraphics[width=\textwidth]{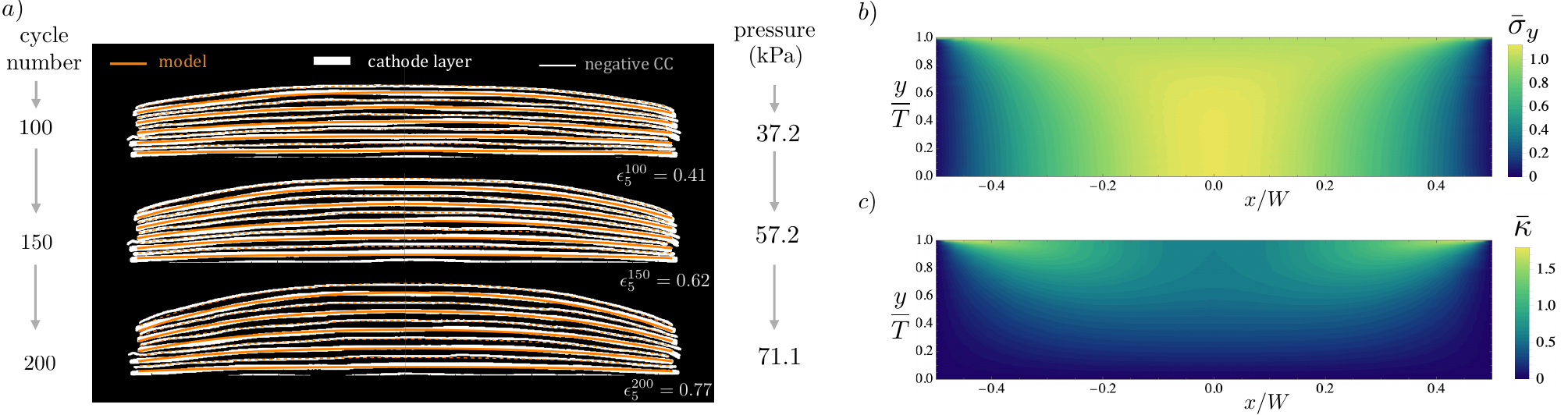}
    \caption{ a) Comparison between the experimental data at different charge cycle and the fitted theoretical model, where orange line represent the theoretical position of the mid-plane of the cathode  bending layers (going though the positive CCs) while dashed lines that of the negative CCs. On the right, the prediction of the pressure. b) The dimensionless stress $\bar{\sigma}$ and bending moment $\bar{M}$ fields predicted by the continuous model. All plots are made with $\gamma=3.21$.}
    \label{fig:final}
\end{figure*}

We now have a theoretical prediction for the shape of the bulging cell for a given parameter $\gamma$ and $\epsilon_n$ (which depend on $\bar{B}$, $K$ and $P$). We fit our model to the experimental data in \cite{Du} at the three different stages of bulging,  during cycle $100$, $150$ and $200$.
We consider only the last (top) cathode layer in the battery, as this is thicker and thus better resolved by the X-ray tomography. After aligning each image and considering only the top half of the cell, we find that the thickness is $T=1.8$ mm (this is the distance between the symmetry plane and the  side of the outermost cathode layer
). We assume that $\gamma$, which depends only on $K$ and $B$, is constant and independent on the cycle number. We fit our model to the data (by minimising the square distance between the theoretical position of the cathode bending layer and the experimental one) to obtain $\gamma$ and three values of the through-cell strain at the different cycle number (shown as a superscript):
$$\gamma=3.21, \quad \epsilon_n^{100}=0.41, \quad \epsilon_n^{150}=0.62, \quad \epsilon_n^{200}=0.77.$$
Our fitted curves are shown in figure \ref{fig:final}a). Crucially, although we have only used the top layer to fit our model to the data, all other layers are very well approximated, suggesting our model is robust. 

In figure  \ref{fig:final}b) we show the rescaled stress and bending moment $M=B {\partial^2 v}/{\partial x^2}$, given by
\[
\bar{\sigma}=\frac{\sigma P}{\hat{K}} =\p{\bar{v}}{\bar{y}}, \quad \text{and} \quad \bar{M}=\frac{T M}{W^2 P}=\frac{1}{\gamma^2}\frac{\partial^2 \bar{v}}{\partial \bar{x}^2}
\]
predicted by the continuous model, showing that the tensile stresses are concentrated in the middle of the structure while the bending moments are concentrated near the boundaries at the top of the system, highlighting regions where mechanical degradation may be largest.

\subsection{Estimate of the pressure and gas production}

We may use our fitted parameters to estimate the pressure within the battery. To do so, we need to estimate the bending stiffness of the bending layers: the cathode layers (including its CC) and the positive current collector, to find $\bar{B}$ (Details in Appendix B). The stiffness of a CC is typically $100$ GPa, their Poisson ratio $\nu_{CC}=0.2$ and their thickness about $10-20 \mu$m, meaning their bending stiffness is roughly $B_{CC}\approx 7 \times 10^{-6}$ Pa m$^2$. Since the cathode layers are not observed to undergo significant material degradation during the bulging process, we assume their stiffness is unaltered and about $1-10$ GPa and the Poisson ratio is $\nu_C=0.2$. From the experimental data, the thickness of the cathode electrode can be as large as $t_C=100 \mu$m. Therefore, a large estimate for the bending stiffness for the cathode layer is $B_C \approx 7 \times 10^{-5}$ Pa m$^2$. Note that, as anticipated, thin current collectors have a significantly lower bending stiffness when compared to cathode bending layers. Thus, the average bending stiffness is $\bar{B}\approx B_C$. The anodes make out for about half of the total thickness of the battery, meaning $\phi\approx0.5$. We can use equation \eqref{eq:gamman} and $\gamma=3.21$ from our fit to the experimental data to find that the stiffness of the substrate and the pressure in the system for the $100^{th}$, $150^{th}$ and $200^{th}$ cycles are:
\begin{equation}
K=92.5\,kPa, \qquad P_{100}=37.2 \,kPa,\qquad
P_{150}=57.2 \,kPa, \qquad P_{200}=71.1 \,kPa.
\end{equation}
 That is, the largest pressure is about an atmosphere --- a more reasonable estimate compared to the naive guess proposed in the introduction. However, this effectively means that the through-cell stiffness in the cell is significantly lower than the pristine reported values. This may be explained in two ways: first, the stiffness of the anode differs when measured under tension or compressions. Typically, the anode softens under tension \cite{uccel2022}. For large strains, this may explain part of the softening observed. Second, in the X-ray tomography data, figure 2 in \cite{Du}, significant damage of the anode is observable, with large pockets of gas forming and altering the porous structure of the electrode. This degradation presumably significantly softens the anode's tensile response. 
 
Finally, we can use our model to estimate the quantity of gas produced in the battery. Since the pressure in the cell remains relatively low, we may assume that the gas within the cell satisfies the ideal gas law:
\begin{equation}
    P V= n_g R \mathcal{T}
\end{equation}
where $P$ is the pressure, $V$ is the volume of gas, $n_g$ represents the number of moles of gas, $R$ is the ideal gas constant and $\mathcal{T}$ the temperature --- which we assume to be constant.

\begin{figure*}[t]
    \centering \includegraphics[width=0.8\textwidth]{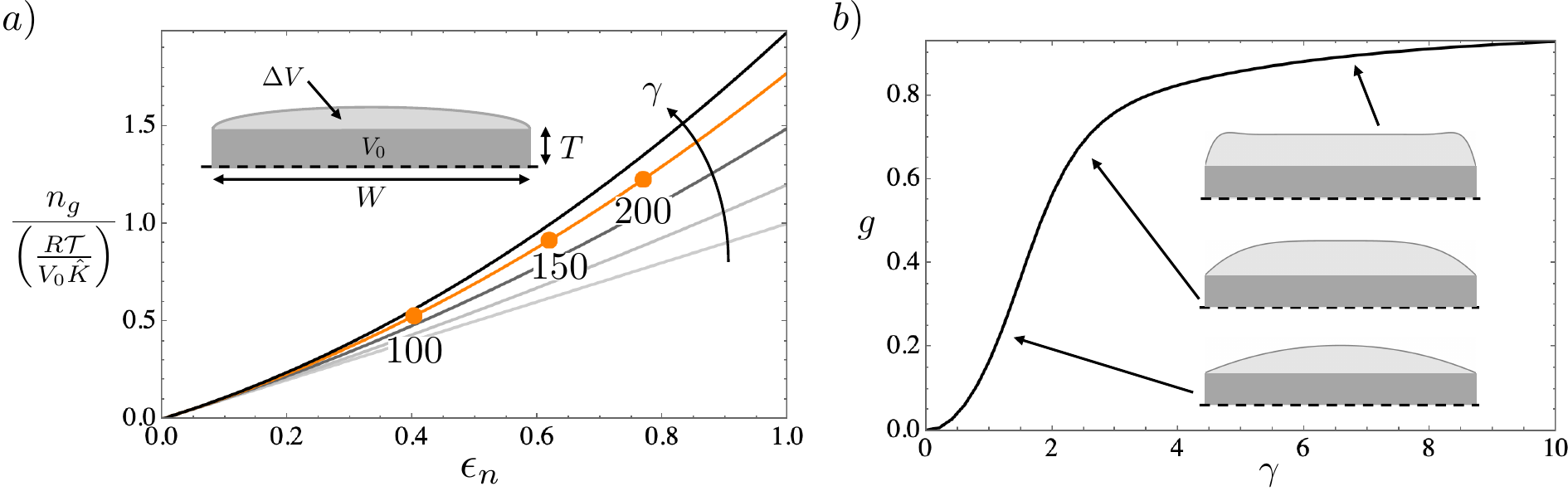}
    \caption{(a) The relationship between the amount of gas $n_g$ and the rescaled pressure $\epsilon_n=P/\hat{K}$ at different values of $\gamma$. In the inset, schematic of the change in volume in the cell. The orange dots represnet the estimated position of the experimental data from \cite{Du} for cycle number $100$, $150$ and $200$.  (b) Shows the $\gamma$ dependence of the relative volume via the function $g(\gamma)$.}
    \label{fig:gas}
\end{figure*}

As the cell expands both the pressure and the volume change. The latter can be estimated from our solution of the shape of the bulge. The change in volume due to expansion is given by
\begin{equation}
    \Delta V=L \int_{-\half W}^{\half W}v(x,T) \mathrm{d}x=V_0 \epsilon_n \int_{-\half}^{\half} \bar{v}(\bar{x},1) \mathrm{d}x
\end{equation}
where $V_0=LWT$ is the pristine volume of the cell. Note that the latter integral only depends on the parameter $\gamma$. Defining $g(\gamma)=\int_{-\half}^{\half} \bar{v}(\bar{x},1;\gamma) \mathrm{d}x$, since $P=\epsilon_n \hat{K}$, we may rewrite the ideal gas law as:
\begin{equation}
\label{eq:gas}
        \frac{n_g}{\left(\frac{V_0 \hat{K}}{R \mathcal{T}}\right)}=\epsilon_n(1+\epsilon_n g(\gamma)).
\end{equation}
In figure \ref{fig:gas}a) we show the evolution of the amount of gas a function of pressure, while in (b) we show a plot of $g(\gamma)$. Crucially, as the rescaled pressure ($\epsilon_n$) increases, the amount of gas grows super-linearly. This observation confirms that estimating the amount of chemical degradation (i.e. the amount of gas formed in the cell) to monitor the SOH of the battery requires more than just knowledge of the internal pressure; it requires, as provided by our model, the relationship between pressure and the volume expansion. Therefore, equation \eqref{eq:gas} offers a direct way to monitor the amount of gas and thus the state of health of the battery from direct observation of its mechanical deformation.

 \section{Conclusion}
Gas generation can lead to the build up of significant pressure in pouch cell batteries causing a large bulging deformation. Measuring the pressure may be a way to control the state of health of the system, but it is not easily done and requires access to the sealed casing. However, information on the pressure may be extracted from the shape of the deformation.

In this paper, we have proposed a model to characterise the shape of the bulging deformation. Guided by the recent detailed experimental images in \cite{Du} of such deformation, we assumed that the anodes are softer and account for all the through-cell strains observed while the cathodes and the current collectors act as stiffer bending sheets. The internal pressure tries to expand the cell in the through-cell direction, but the casing prevents the sides from expanding, leading most of the deformation to occur in the middle, in the form of a bulge. By assuming that the layers are many and thin, consistent with the structure of most pouch cells, we derived a homogenised theory which leads to a closed form expression for the shape of the bulge, equation \eqref{eq:vbar}. The shape depends on two dimensionless numbers, $\gamma=(W/l_n)$ --- which measures the relative importance of the width $W$ of the system compared to the typical length over which bending deformations persist into the bulk of the battery, $l_n$ --- and the strain induced by the pressure, $\epsilon_n=P/K$. 

We have shown that our homogenised theory works well even when the number of layers is moderate, $n=5$, provided $\gamma>3$, and fit our solution to the experimental data in \cite{Du} to find values of $\gamma$ and through-cell strain $\epsilon_n$ at the $100^{th}$, $150^{th}$ and $200^{th}$ cycle. Then, making a reasonable assumption about the value of the bending stiffness of the cathodes, we have estimated that the pressure in the system during bulging to be of the order of $30-70$ kPa.

We have used our result to predict the quantity of gas produced in a battery as a function of the pressure and shape of the bulge, equation \eqref{eq:gas}. Since gas production is a key indicator of the state of health of a cell, our model offers a novel way to control the SOH of the battery without breaking the sealed casing.

Finally, we comment on some of the assumptions made to derive our model. We have chosen to model our battery as a 2D layered structure. Our argument to justify this choice was that the length $L$ is larger than the width $W$ (more than twice), so that the deformation is observed to be independent of the $z$ direction in the bulk, although a boundary layer near $z=\pm L/2$ exists, since, also there, the casing clamps the structure. As we have seen, the typical length-scale over which a deformation on the side of the structure persists is $l_n=W/\gamma$ which, for our battery, is $l_n=7.07mm$. This is almost an order of magnitude smaller than the length $L$. Indeed, if we were to solve our problem along $z$ (assuming plane strain in the $y$ direction) we would find that $\gamma \approx 7$ and, as shown in figure \ref{fig:NDmodel}, the effects of the boundary  would have substantially decayed into the bulk, meaning our 2D model is self consistent.

\newpage

\begin{appendices}

\section{Appendix A: Energy stored in a thin anode layer}
\label{sec:appendixA}

Here, we show that, when $T \ll L, W$, the deformation in an anode electrode (between a cathode and a CC) is well approximated by assuming the anode behaves as a Winkler foundation. 
Without loss of generality, we only consider the anode between the $i$ and $i+1$ layers, with $y \in [y_i+\half t_{CC},y_{i+1}- t_C-\half t_{CC} ]$, where $y_i$ is the position of the $i^{th}$ negative CC and $y_{i+1}$ that of the mid-plane of the negative CC between cathode layers. As the system bulges, the anode is deformed, with a (plane-strain) deformation given by the displacements $\vec{u}(x,y)=(u(x,y),v(x,y),0)$.  

This displacement-field generates a strain:
\begin{equation}
    \vec{\varepsilon}=\half(\nabla \vec{u}+\nabla \vec{u}^T)
\end{equation}
and a stress 
\begin{equation}
    \vec{\sigma}=\frac{E_A}{1+\nu_A}\left(\vec{\varepsilon}+\frac{\nu_A}{1-2\nu_A}\Tr{\vec{\varepsilon}} \right)
\end{equation}
where $E_A$ and $\nu_A$ are the Young's modulus and Poisson ratio of the anode material. 

The anode material must satisfy the equilibrium condition 
\begin{equation}
\label{eq:div}
    \nabla \cdot \vec{\sigma}=0.
\end{equation}
Note that, although $v_i(x)$ is the mid-plane of the cathode layers, since their thickness remain unchanged and rotations are small ($v_i'(x)\ll 1$), displacing the cathode by $v_i$ means displacing the interface between cathode and anode by $v_i(x)$, and hence 
\begin{equation}
\label{eqA:bc}
    v(x,y_i+\half t_{CC})=v_i(x), \quad \text{and} \quad v(x,y_{i+1}-t_C-\half t_{CC})=v_i+\Delta v(x)=v_{i+1}.
\end{equation}

Finally, since the current collectors are inextensible, the system must also satisfy the inextensibility constraint 
\begin{equation}
\label{eqA:const}
    \p{v}{x}^2+\left(1+\p{u}{x}\right)^2=1,
\end{equation}
at $y=y_i$ and $y=y_{i+1}$.
We need to solve the equilibrium condition \eqref{eq:div} with boundary condition \eqref{eqA:bc} and constraint \eqref{eqA:const}.
We let $T^*$ to be the typical thickness of the cell and $W^*$ the typical length. Let $\delta = T^*/W^* \ll 1$ and let the typical stiffness of the anode be $E^*$, so that we can rescale our quantities as
 \begin{align}
     W&=W^* \tilde{W}, \quad    T= T^*  \tilde{T}, \quad   t_{CC} = T^* \tilde{t}_{CC} ,\quad   t_C = T^* \tilde{t}_C ,  \quad   t_A= T^* \tilde{t}_A,\\
     x &= W^* \tilde{x},  \quad   y = T^* \tilde{y}, \quad
     u= W^* \tilde{u},  \quad   v= W^* \tilde{v},  \quad   v_i=T^* \tilde{v}_i ,  \quad   \sigma= E^* \tilde{\sigma},
 \end{align}
We now expand our equilibrium equation and constraint in small $\delta$ so that a generic dependent variable $\phi$ becomes:
\begin{equation}
    \phi(x,y)=\phi^{(0)}(x,y)+\phi^{(1)}(x,y)\delta+\phi^{(2)}(x,y)\delta^2 +\dots
\end{equation}
At leading order, we have that
$$\frac{\partial^2 \tilde{u}^{(0)}}{\partial \tilde{y}^2}=\frac{\partial^2 \tilde{v}^{(1)}}{\partial \tilde{y}^2}=0.$$ 
The inextensibility constraint, equation \eqref{eqA:const}, leads to
$$\p{\tilde{u}^{(0)}}{\tilde{x}}=0$$ 
at $\tilde{y}=\tilde{y}_i$ and $\tilde{y}=\tilde{y}_{i+1}$.
The boundary conditions require $\tilde{v}(\tilde{x},\tilde{y}_i)=\tilde{v}_i(x)$ and $\tilde{v}(\tilde{x},\tilde{y}_i+\half \tilde{t}_{CC}+\tilde{t}_A)=\tilde{v}_i(x)+\Delta \tilde{v}_i(x)$, where $\Delta \tilde{v}_i(x)=\tilde{v}_{i+1}(x)-\tilde{v}_i(x)$. Putting everything together, we find the general solution:
$$\tilde{u}^{(0)}=\tilde{u}_0+\tau (\tilde{y}-\tilde{y}_i +\half \tilde{t}_{CC}) $$
\begin{equation}
\label{eqA:v}
    \tilde{v}^{(1)}(x,y)=\tilde{v}_i(x)+\Delta \tilde{v}_i(x) (\tilde{y}-\tilde{y}_i+\tilde{t}_{CC})/\tilde{t}_A,
\end{equation}
where $\tilde{u}_0$ and $\tau$ are constants, one representing solid body displacement along the $x$ direction (which we set to $0$), the other representing shear in the $x$-$y$ plane. 

We can use our leading order fields to write down the leading order term in the (non-dimensional)  energy density, $\tilde{\mathcal{W}}= \half \Tr{\tilde{\sigma} \cdot \tilde{\varepsilon}}$, given by
$$\tilde{\mathcal{W}} = \frac{E_A}{2(1+\nu_A)}\left(\frac{\tau^2}{2}+\frac{(1-\nu_A)\Delta v_i(x)^2}{t_A^2(1-2 \nu_A)} \right)$$
We can minimise this energy with respect to shear $\tau$ and find that $\tau=0$. This means that our leading order displacements are only in the through-cell direction, and $u^{(0)}=0$ --- the anode behaves as a Winkler foundation. 

We can write down the energy per unit length along the $x$ direction. In dimensional form, it is given by $\int_{-L/2}^{L/2}\int_{y_i+\half t_{CC}}^{y_i+\half t_{CC}+t_A} W dy dz$, and reads:
\begin{equation}
 U_{el}= \int_{W/2}^{-W/2} \half \delta L t_A K  \left( \frac{\Delta v(x)}{t_A}\right)^2 dx
\end{equation}
where $K=\frac{ E_A (1-\nu_A)}{(1+\nu_A)(1-2 \nu_A)}.$
This is the energy stored in a Winkler foundation.

\section{Appendix B: Bending stiffness of a cathode plus current collector layer.}
\label{appendix:B}

The bending stiffness of a CC sheet of thickness $t_{CC}$ (in plane strain) is given by:
\begin{equation}
    B_{CC}=\frac{E_{CC}}{(1-\nu_{CC}^2)}\int_{-\half t_{CC}}^{\half t_{CC}} y^2 dy = \frac{E_{CC}}{12(1-\nu_{CC}^2)} t_{CC}^3.
\end{equation}
If we consider also the two cathode layers (on either side of a CC), as in the case of a cathode bending sheet, we need to add to the contribution from straining the cathodes, so that  
\begin{align}
B_{C}&=B_{CC}+\frac{2 E_{C}}{(1-\nu_{C}^2)}\int_{\half t_{CC}}^{\half t_{CC}+t_C} y^2 dy \\
&= \frac{E_{CC}}{12(1-\nu_{CC}^2)} t_{CC}^3-\frac{E_{C}}{12(1-\nu_{C}^2)} t_{CC}^3 + \frac{2 E_{C}}{3(1-\nu_{CC}^2)} (\half t_{CC}+t_{C})^3.
\end{align}

\section{Appendix C: Numerical simulations of the bulging battery}
\label{sec:app_num}

\begin{figure}[h]
    \centering \includegraphics[width=0.5\textwidth]{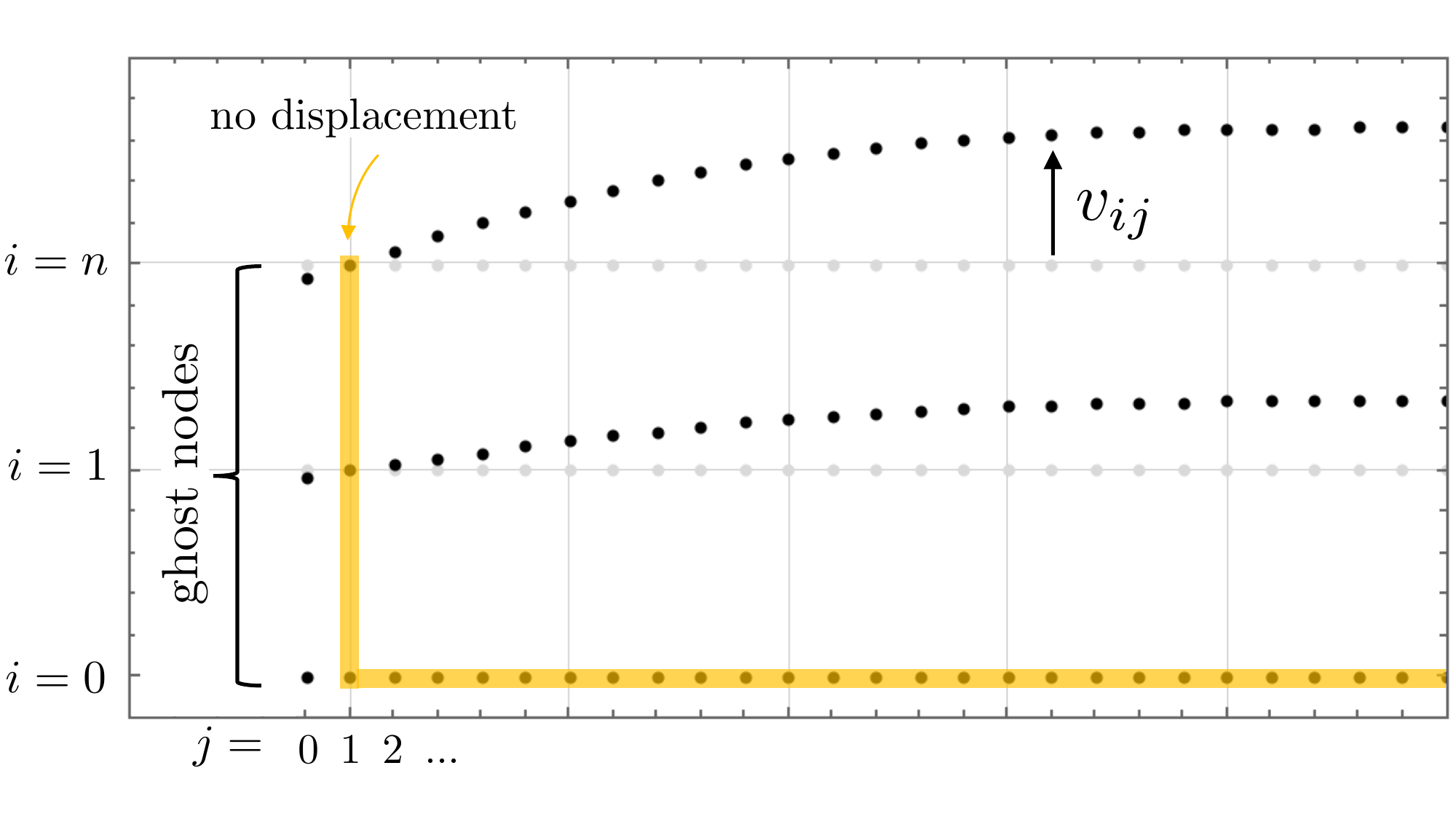}
    \caption{ Schematics of the discretisation used to minimise the energy in equation \eqref{eqn:energy_tot_n}.}
    \label{fig:finalenergy}
\end{figure}

We can find the shape of the bulging system from the energy in equation \eqref{eqn:energy_tot_n}. Instead of minimising the energy and solving a system of $n$ coupled differential equations for the $v_i$s, we discretise the system and minimise the energy numerically. We discretise the $x$ axis in $n_j$ segments with $x_j=-W/2+W (j-1)/n_j$ with $j$ ranging from $1$ to $n_j+1$. In our numerics, we chose $n_j=200$ as this proved sufficient for convergence of the results. 

To account for second derivatives (bending) we include two ghost nodes, so that $j \in \{0, n_j+2\}$. We then can write $v_{ij}=v_i(x_j)$. 
We discretise the curvature as:
\begin{equation}
    \frac{\partial^2 v_i}{\partial x^2}=\frac{v_{i(j-1)}+v_{i(j+1)}-2v_{ij}}{(x_{(j+1)}-x_j)^2}.
\end{equation}
Then, the energy in \eqref{eqn:energy_tot_n} can be written as a function of all the $v_{ij}s$: 
   \begin{equation}
    \label{eqn:energy_tot_disc}
    \tilde{U}_{tot}(v_{ij})=\sum_{j=1}^{n_j}\sum_{i=1}^{n}\left( \half t_A K \left(\frac{(v_{ij}-v_{(i-1)j})}{t_A}\right)^2 +\half (\bar{B}+(-1)^{i-1} \Delta B )\left(\frac{v_{i(j-1)}+v_{i(j+1)}-2v_{ij}}{dx^2} \right)^2 - P(v_{ij}-v_{(i-1)j})\right).
    \end{equation}
We can minimise this energy with respect to $v_{ij}$ subject to the clamped boundary condition $v_{i0}=v_{i(n_j+2)}=0$ and $v_{0j}=0$ using the function \emph{NMinimise} on \emph{Wolfram Mathematica}. The results are shown in figure \ref{fig:NDmodel}.
\end{appendices}

\bibliographystyle{apsrev4-2}
\bibliography{biblio}  

\end{document}